\documentclass[%
preprint%
 ,secnumarabic%
,amssymb, amsmath,nobibnotes,aps]{revtex4}
\usepackage{epsfig}%
\usepackage{graphicx}%
\usepackage[lofdepth,lotdepth]{subfig}%
\expandafter\ifx\csname package@font\endcsname\relax\else
 \expandafter\expandafter
 \expandafter\usepackage
 \expandafter\expandafter
 \expandafter{\csname package@font\endcsname}%
\fi
\begin{document}

\title{Constraining CSL strength parameter $\lambda$ from standard cosmology and spectral distortions of CMBR}%

\author{Kinjalk Lochan}
\email{kinjalk@tifr.res.in}
\author{Suratna Das}%
\email{suratna@tifr.res.in}
\affiliation{Tata Institute of Fundamental Research, Mumbai 400005, India}
\author{Angelo Bassi}%
\email{bassi@ts.infn.it}
\affiliation{Department of Physics,
University of Trieste, Strada Costiera 11, 34151 Trieste, Italy.
\\ IIstituto Nazionale di Fisica Nucleare, Trieste Section, Via Valerio 2, 34127 Trieste,
Italy.}
\def\be{\begin{equation}}
\def\ee{\end{equation}}
\def\al{\alpha}
\def\bea{\begin{eqnarray}}
\def\eea{\end{eqnarray}}

\begin{abstract}
Models of spontaneous wave function collapse modify the linear
Schr\"{o}dinger equation of standard Quantum Mechanics by adding
stochastic non-linear terms to it. The aim of such models is to
describe the quantum (linear) nature of microsystems along with the
classical nature (violation of superposition principle) of macroscopic
ones. The addition of such non-linear terms in the Schr\"{o}dinger
equation leads to non-conservation of energy of the system under
consideration. Thus, a striking feature of collapse models is to heat
non-relativistic particles with a constant rate. If such a process is
physical, then it has the ability to perturb the well-understood
thermal history of the universe. In this article we will try to
investigate the impacts of such heating terms, according to the
Continuous Spontaneous Localization (CSL) model, on standard evolution
of non-relativistic matter and on the formation of CMBR. We will also
put constraints on the CSL collapse rate $\lambda$ by considering that
the standard evolution of non-relativistic matter is not hampered and
the observed precise blackbody spectrum of CMBR would not get
distorted (in the form of $\mu-$type and $y-$type distortions) so as
to violate the observed bounds.
\end{abstract}
\maketitle
\section{Introduction}

Models of spontaneous wave function collapse \cite{Bassi,
  Bassi:2012bg} aim to unify the dynamics of microscopic and
macroscopic systems in order to answer the long prevailing question of
Quantum ``Measurement problem''. The unification of microscopic
(superposition of states) and macroscopic (violation of superposition
principle) dynamics is accomplished by modifying the Schr\"{o}dinger
equation through adding non-linear stochastic terms. The non-linear
terms in the modified Schr\"{o}dinger equation ensures the breakdown
of superposition principle at the macroscopic level and the stochastic
nature of such dynamics indicates that the outcome of measurements
would be probabilistic. The added non-linear terms act as
amplification mechanism to ensure that these modifications have
negligible impacts on microscopic system but are very efficient of
localization for macroscopic ones. It is also important to note that
the added stochastic terms also respect causality of the dynamics as
deterministic non-linear evolution of Schr$\ddot{\rm o}$dinger
equation leads to violation of relativity as shown in
\cite{Gisin}. Among the many attempts of constructing such a modified
dynamics, Quantum Mechanics of Spontaneous Localization model (QMSL,
later known as GRW model after the name of the authors
\cite{Ghirardi}), Quantum Mechanics with Universal Position
Localization model (QMUPL \cite{Diosi, Bassi:qmupl}) and Continuous
Spontaneous Localization (CSL \cite{Ghirardi:1989cn, PS}) model are
worth noting. In some models it is considered that localization of
wave-packets is a consequence of gravitational effects
\cite{Karolyhazy, Diosi-gr, penrose}. In this article we will
concentrate only on CSL model and analyze the signatures of such a
model while applying it in cosmology.

According to the present understanding of the CSL model, a theoretical
origin of the free parameters introduced in the scheme, i.e. the
collapse rate $\lambda$ and the width of the localization $r_c$, are
yet to be determined. Hence values of such free parameters have to be
obtained phenomenologically. Several proposed bounds on the free
parameter $\lambda$ (considering $r_c\approx10^{-5}$ cm) can be found
in \cite{adler-science, Bassi:2012bg}.
  
One distinguishing feature of the CSL model, common to most collapse
models, is that due to the presence of non-linear stochastic terms in
the Schr$\ddot{\rm o}$dinger equation the total energy of the system
does not remain constant and the non-relativistic massive particles
within a system gain energy with a constant rate proportional to the
CSL parameter $\lambda$. Such heating effects are interesting to study
for systems which are generally in thermal equilibrium. Study of
atomic and nuclear systems \cite{PS} which demands the
non-dissociation of cosmic hydrogen by CSL heating obtains an upper
bound as $\lambda<1$ s$^{-1}$, whereas studies of proton decay
considering CSL models \cite{Adler} also lead to similar
bounds. Comparing with the experimental bounds on photon emission from
Germanium one can constrain the CSL parameter as $\lambda <10^{-11}$
s$^{-1}$. Furthermore, in a cosmological scenario heating of protons
through CSL mechanism over universe's life-time \cite{Bassi, Ghirardi,
  Adler} and the rise of CMBR temperature due to interactions of CMBR
photons with these heated protons \cite{Adler} have been studied to
suggest a bound as $\lambda <10^{-5}$ s$^{-1}$. Also thermal
equilibrium of Inter Galactic Medium (IGM) \cite{Adler} has been
studied to put an upper bound as $\lambda <10^{-8}$ s$^{-1}$.

However, as the thermal evolution of our universe is well studied and
constrained observationally, there can be other cosmological
scenarios, apart from the evolution of IGM, where the imprints of the
anomalous heating of CSL model can be studied. One such scenario,
which we will exploit in the present article, is the thermodynamic
equilibrium of matter and radiation before the formation of
CMBR. Injection of energy during such epochs perturbs the
thermodynamic equilibrium of matter and radiation in the cosmic plasma
which leads to distortions in the well-measured blackbody spectrum of
CMBR. Such distortions are stringently constrained by
observations. Here we will address two types of distortions of CMBR :
the $\mu-$distortion, which results from energy releases during the
redshift span $2\times10^6>z>5\times10^4$, and the $y-$distortion,
which comes from energy releases during the redshift span
$1100<z<5\times10^4$. $\mu-$distortion yields a non-vanishing,
frequency-dependent small chemical potential of the cosmic photons
which leads to a Bose-Einstein distribution in the high-frequency
regime of the spectrum rather than the pure Planckian spectrum of a
blackbody \cite{Sunyaev:1970er}. Observations of COBE/FIRAS pointed
out that such a distortion should be very small and puts an upper
bound on such distortions as $\mu\lesssim 9\times10^{-5}$ with $95\%$
confidence \cite{cobe}, whereas an upcoming experiment PIXIE can probe
such distortions up to $\mu\sim5\times10^{-8}$ \cite{Kogut:2011xw}. On
the other hand, the $y-$distortion of the spectrum is characterized by
the lower density of photons in the low-frequency regime and increment
of photon number in the high-frequencies with respect to a standard
Planckian blackbody spectrum \cite{Zeldovich:1969ff}. COBE/FIRAS
observations put an upper bound on observed $y-$parameter as
$y\leq1.5\times10^{-5}$ with $95\%$ confidence \cite{cobe}, whereas
PIXIE can put an upper bound on such distortion as $y\leq10^{-8}$
\cite{Kogut:2011xw}.

In this present article we will explore the possibilities of
generating spectral distortions of CMBR due to CSL heating of
non-relativistic particles. Before any such endeavor, one should
confirm that evolution of non-relativistic matter during the radiation
and matter dominated era should not get affected by such anomalous
heating of CSL and they evolve according to the standard
cosmology. Then we note here that as such an anomalous CSL heating of
non-relativistic matter can disturb the thermodynamic equilibrium of
matter and radiation before CMBR formation, it can lead to spectral
distortions of the CMBR where the amount of distortions will depend
upon the strength of the heating of the particles. Thus by quantifying
the amount of distortions such CSL heating can generate, one can put
bounds on the free parameter $\lambda$ of the theory.

However, in the present article we will not consider the effects of
CSL heating on the evolution of relativistic matter including
radiation. It has been a challenge to consistently formulate any
relativistic generalization of the collapse models such as CSL and
various attempts have been made in that direction. For a brief review
of the attempts made to make a relativistic generalization of
Spontaneous Collapse models one may refer to
\cite{Bassi:2012bg}. Thus, here we will assume that the relativistic
matter evolve according to the standard cosmology. Any properly
developed relativistic collapse model can, in principle, leave its
imprint on CMBR and will be worth studying in future.

According to standard Friedmann-Robertson-Walker (FRW) cosmology at
$z_i\approx2\times10^6$ the temperature of the cosmic soup is
$T_i\approx470$ eV which is much lower than the rest mass energy of
electrons which is $0.5$ MeV. Thus the electrons and protons present
in the cosmic soup during the concerned epochs
i.e. $2\times10^6>z>1100$ can be treated as non-relativistic
particles. Furthermore, if one tracks the density profile
$n(z)=n_0(1+z)^3$, where $n_0$ is the present number density of
baryons today, then one obtains the number of baryons per unit
correlation cell of size $r_c^3\sim10^{-21}$ m$^{3}$ (prescribed in
CSL model) to be smaller than unity at $z_i\approx10^6$ which falls
off even further with the expansion of the universe. This indicates
that considering the electrons and protons as free particles during
such epochs is a good approximation. Hence the above discussion shows
that during such epochs CSL model is applicable considering the
massive particles in the cosmic soup to be free and
non-relativistic. This enforces one to seek for possible spectral
distortions in the CMBR generated due to CSL heating and to put bounds
on $\lambda$ by quantifying the amount of distortions that can be
yielded by such heating.

We have organized the present article in the following manner.  In
Sec.~(\ref{CSL}) we will briefly review the CSL model and estimate the
rate of heating of non-relativistic particles due to presence of
non-linear stochastic terms in the Schr\"{o}dinger equation in the CSL
scenario. Here, we will discuss two scenarios of CSL model : firstly
the generic case where the strength parameter $\lambda$ is independent
of mass and a variant scenario where $\lambda$ is dependent on the
mass of the non-relativistic particles under consideration. We have
mentioned before that one should ensure that CSL heating would not
disturb the standard evolution of the non-relativistic matter during
radiation and matter dominated era. Sec.~(\ref{stand-cosmo}) is
devoted to such discussions and to constrain the strength parameter
$\lambda$ by demanding the standard cosmological evolution of
non-relativistic matter throughout. In Sec.~(\ref{mu-dist}) we will
first give a brief review of $\mu-$distortion and then we will put
bounds on both the scenarios of CSL model discussed
above. Sec.~(\ref{y-dist}) is focused to discuss the $y-$distortion
that CSL heating of non-relativistic particles can generate and
constraining the CSL parameter from such distortion of CMBR
spectrum. In the penultimate section (Sec.~(\ref{photon})) we will
briefly consider the case where spontaneous photon emission is taken
into account within the arena of CSL model and show that such feeble
emission of photon in the high-frequency regime would be insufficient
to generate any considerable distortion in the CMBR spectrum and thus
would not be an important feature to constrain the strength parameter
$\lambda$. In the last section we will discuss all the bounds we will
obtain throughout and then conclude.

\section{Reviewing CSL model and explaining the heating of particles}
\label{CSL}

In this section we will briefly review the CSL mechanism and how it
leads to heating of non-relativistic particles with a constant rate.
We will initially describe the GRW model of spontaneous collapse
\cite{Ghirardi} where randomly acting non-linear terms are added to
the Schr$\ddot{\rm o}$dinger equation, based on the assumption that
the constituents of a system suddenly collapse to a localized
wave-function with an appropriate rate. Concepts of such models are
easy to conceive and the main result for energy gain by the
non-relativistic particles are the same in both GRW and CSL models
which makes the GRW model worth describing at this point. In
literature, at later times GRW model has been extended to CSL model by
appropriately introducing stochastic terms in the Schr$\ddot{\rm
  o}$dinger equation combined with the theory of continuous Markov
processes in Hilbert space. We will discuss the CSL model in the later
subsection.

\subsection{GRW model of spontaneous wave function collapse}

A deviant of quantum mechanics, called the GRW model \cite{Ghirardi},
allows for spontaneous localization of a particle with a mean rate
$\lambda$, where the evolution of the system between two successive
localizations is according to the standard Schr$\ddot{\rm o}$dinger
equation. According to this theory a localization operator,
\begin{eqnarray}
L_{\mathbf x}^i=\left(\frac{\alpha}{\pi}\right)^{\frac34}e^{-\frac\alpha2({\mathbf q}_i-{\mathbf x})^2},
\end{eqnarray}
localizes a wave-function $|\psi\rangle$ in space (at point $\mathbf x$)
yielding $|\psi^i_{\mathbf x}\rangle/||\,|\psi^i_{\mathbf
    x}\rangle\,||$, where $|\psi^i_{\mathbf x}\rangle \equiv L_{\mathbf x}^i |\psi\rangle$ is the
wave-function of the $i^{\rm th}$ particle localized at $\mathbf
x$. Here $1/\sqrt{\alpha}\equiv r_c$ quantifies the accuracy of the
localization at position ${\mathbf q}_i$. Due to the stochasticity of the spontaneous
localization, a system in a pure state, which consists of a single
particle, is transformed into a statistical mixture :
\begin{eqnarray}
|\psi\rangle\langle\psi|\rightarrow \int d^3x L_{\mathbf x}^i|\psi\rangle\langle\psi|L_{\mathbf x}^i\equiv T\left[|\psi\rangle\langle\psi|\right].
\end{eqnarray}
Analogously, if the system is initially in a statistical mixture,
given by the operator $\rho$, then the localization of the system is
determined by $T[\rho]$ with a probability $\lambda$. Hence the
statistical state $\rho$ after time-interval $dt$ can be obtained as
\begin{eqnarray}
\rho(t+dt)=(1-\lambda dt)\left[\rho(t)-\frac{i}{\hbar}[H,\rho(t)]dt\right]+\lambda dt T[\rho(t)],
\end{eqnarray}
where the first term on R.H.S. indicates that the system evolves
according to Schr$\ddot{\rm o}$dinger equation if it is not localized
within the interval $dt$. Thus the evolution of the system,
represented by operator $\rho$, will be according to the {\it master
equation} :
\begin{eqnarray}
\frac{d}{dt}\rho(t)=-\frac{i}{\hbar}[H,\rho(t)]-\lambda\left(\rho(t)- T[\rho(t)]\right).
\label{rho-qmsl}
\end{eqnarray}
A simple example of free particle Hamiltonian in one dimension has
been considered in \cite{Bassi, Ghirardi} where the solution of the
master equation in coordinate space has been obtained as 
\begin{eqnarray}
\left\langle q^\prime\left|\rho(t)\right|q^{\prime\prime}\right\rangle=\frac{1}{2\pi\hbar}\int_{-\infty}^{+\infty}dk\int_{-\infty}^{+\infty}dy e^{-\frac{i}{\hbar}ky}F(k,q^\prime-q^{\prime\prime},t)\left\langle q^\prime+y\left|\rho_{\rm S}(t)\right|q^{\prime\prime}+y\right\rangle,
\end{eqnarray}
where $\left\langle q^\prime\left|\rho_{\rm
  S}(t)\right|q^{\prime\prime}\right\rangle$ is the solution of the
pure Schr$\ddot{\rm o}$dinger equation and the factor
\begin{eqnarray}
F(q,k,t)=e^{-\lambda t+\lambda\int_0^td\tau e^{-\frac{\alpha}{4}(q-k\tau/m)^2}}
\end{eqnarray}
encapsulates all the dynamics of localization. $\lambda\rightarrow
0$ yields $F(q,k,t)=1$ which shows in this limit $\left\langle
q^\prime\left|\rho_{\rm
  S}(t)\right|q^{\prime\prime}\right\rangle=\left\langle
q^\prime\left|\rho(t)\right|q^{\prime\prime}\right\rangle$, as
expected. The dynamical evolution of this free particle system yields
a spread in the momentum as \cite{Ghirardi}
\begin{eqnarray}
\langle\hat{p}^2\rangle\equiv\rm{tr}[\hat{p}^2\rho(t)]=\langle\hat{p}^2\rangle_{\rm S}+\frac{3\alpha\lambda\hbar^2}{2}t,
\end{eqnarray}
which can be derived using Eq.~(\ref{rho-qmsl}). Here, $\langle\hat{p}^2\rangle_{\rm S}$ is the conserved momentum for the free
Schr\"odinger evolution. Thus for a
non-relativistic particle with energy $E=\frac{p^2}{2m}$ the spread in
energy will be
\begin{eqnarray}
\langle E\rangle=\langle E\rangle_{\rm S}+\frac{3\alpha\lambda\hbar^2}{4m}t,
\end{eqnarray}
which shows non-conservation of energy of the system with
time. Therefore, one can infer that due to spontaneous localization of
particles in a system a non-relativistic particle gains energy with a
rate
\begin{eqnarray}
\frac{\partial E}{\partial t}=\frac{3\alpha\lambda\hbar^2}{4m}.
\label{inc-rate}
\end{eqnarray}
This heating rate has been obtained for one non-relativistic
particle. 
To investigate effects of such spontaneous localizations in
cosmology one has to obtain the total heating rate for the cosmic
plasma containing non-relativistic electrons and protons.
\subsection{Continuous Spontaneous Localization (CSL) model}

Out of many spontaneous collapse models, the one that is commonly used
in physical applications is the Continuous Spontaneous Localization
(CSL) model which generalizes the original GRW model \cite{Ghirardi}
to systems of identical particles. In the CSL model the collapse of
the wave-function happens continuously in time in contrast to the GRW
model where the collapse of wave-function happens discretely. The
parameters introduced in this model, in a similar way as had been done
in GRW model, are the coupling constant $\gamma$ and the correlation
length $r_c$ (which is conventionally taken as $10^{-5}$ cm).  The
modified Schr\"{o}dinger equation takes the form in CSL model
\cite{Bassi}
\begin{eqnarray}
|d\psi_t\rangle&=&
\left[-\frac{i}{\hbar}Hdt+\sqrt{\gamma}\int d^3x( N({\bf x})-\langle N({\bf x})\rangle_t)dW_t({\bf x})\right.\nonumber\\
&&\left.-\frac{\gamma}{2}\int d^3x( N({\bf x})-
\langle N({\bf x})\rangle_t)^2dt\right]|\psi_t\rangle. 
\end{eqnarray}
Here the operator $N({\bf x})$ is an averaged density operator
given as
\begin{eqnarray}
N(\mathbf x)=\sum_s\int d^3yg(\mathbf{y}-\mathbf{x})a^\dagger(\mathbf{y},s)a(\mathbf{y},s),
\end{eqnarray}
where the sum is over various particle species $s$, with
$a^{\dagger}_s({\bf y})a_s({\bf y})$ being the number density operator
of species $s$ and $W_t({\bf x})$ is the family of standard Wiener
process for each point in space.  Moreover, $g(\mathbf x)$ is a
spherically symmetric, positive, real function peaked around
$\mathbf{x}=0$ with the normalization $\int d^3x g(\mathbf x)=1$ and
can be written as
\begin{eqnarray}
g({\bf x})=\frac{1}{(2\pi r_c^2)^{3/2}}e^{-{\bf x}^2/r_c^2}.
\end{eqnarray}
The collapse rate $\lambda$ of GRW model is related to that of the
CSL parameter $\gamma$ as
\begin{equation}
 \lambda=\frac{\gamma}{8\pi^{3/2}r_c^3}.
\end{equation}
The corresponding master equation in
CSL model as for Eq.~(\ref{rho-qmsl}) in GRW is given by \cite{Bassi}:
\begin{eqnarray}
\frac{d\rho(t)}{dt}=\frac{i}{\hbar}[H,\rho(t)]+\gamma\int d^3x N(\mathbf{x})\rho(t)N(\mathbf{x})-\frac\gamma2\int d^3x \left\{N^2(\mathbf{x}),\rho(t)\right\}.
\end{eqnarray}
It is to be noted from this relation that $\lambda$ of GRW model is
directly proportional to the parameter $\gamma$ of CSL model apart
from some constant factors. In such a case a bound on $\lambda$ will
directly indicate a bound on $\gamma$. Hence we will consider bounds
in $\lambda$ for further discussions and those can be converted into
bounds on $\gamma$.

In a generic CSL model scenario the introduced parameter $\gamma$ 
(or, equivalently, $\lambda$) is
independent of the mass of the constituent particles. In literature, a
variant of such a model is also discussed where the parameter
$\lambda$ is mass-dependent as
\begin{equation}
 \lambda(m)=\lambda_0\left(\frac{m}{m_N}\right)^{\beta},
\end{equation}
where $m_N$ is the mass of a nucleon. Such a model is motivated from
the feature that in this scenario the collapse rate is different for
different species bearing different masses \cite{PS}. We provide a
justification for such a choice. The calculations of the reduction
rate of wave-functions in \cite{Bassi} shows that the off-diagonal
elements of coordinate space density matrix for a single nucleon
approach zero exponentially with a reduction rate $\Gamma_R$ given by
\begin{eqnarray}
\Gamma_R = \lambda\left(1-e^{-{\bf x}^2/4 r_c^2}\right),
\end{eqnarray}
which for $|{\bf x}|>r_c$ becomes
\begin{equation}
 \Gamma_R \approx \lambda.
\end{equation}
For $n$ nucleons within a radius smaller than the correlation length,
in a superposition of states with distance larger than $r_c$, this
rate is multiplied by $n^2$ to yield
\begin{eqnarray}
\Gamma_R \simeq	 n^2\lambda.
\end{eqnarray}
In mass-proportional CSL model one wants the collapse rate of a
massive single particle composed of $n$ fundamental particles to be
the same as the collapse rate of $n$ single fundamental particles. This
is true when the strength parameter $\lambda$ is quadratically
proportional to the mass of the constituent particle i.e. $\beta=2$.
In that case the reduction rate of the system of $n$ fundamental
particles becomes
\begin{eqnarray}
\Gamma_R = \lambda_0 n^2\left(\frac{m}{m_N}\right)^2.
\end{eqnarray}
Thus in this article, along with the mass-independent CSL model with
strength parameter $\lambda$, we will also analyze the mass-dependent
CSL model where the strength parameter $\lambda_0$ is for the
quadratically mass-dependent case and compare the bounds obtained in
both the cases.

It has been shown in \cite{PS} that due to the presence of non-linear
terms in the Schr\"{o}dinger equation in the CSL model the energy will
not be conserved (as has also been discussed for GRW model). Since, as
discussed in the introduction, the number of baryons in the universe
per unit correlation cell of size $r_c^3\sim10^{-21}$ m$^{3}$ is
smaller than unity, for the considered times, the effect of the
identity of particles can be neglected and the rate energy increase in
such a scenario is similar to that in the GRW model which is given in
Eq.~(\ref{inc-rate}). For CSL model the non-conservation of energy is
quantified as \cite{PS}
\begin{equation}
\langle E\rangle=\frac{3\lambda \alpha \hbar^2}{4 m}t.
\end{equation}
Such a heating rate has been obtained for one non-relativistic
particle. To investigate effects of such spontaneous localization in
cosmology one has to obtain the total heating rate of the cosmic
plasma containing non-relativistic electrons and protons. Given the
number density $n_{s}$ for each species $(s=e,p)$ the total energy
density gain by each species can be straightforwardly obtained from
the above equation as
\begin{eqnarray}
\left.\frac{\partial \varepsilon_s}{\partial t}\right.=\frac{3\alpha\lambda\hbar^2}{4m_{s}}n_{s}\, ,
\label{heating-rate}
\end{eqnarray}
taking into account that during the epochs one is interested in, these
particles are non-relativistic and behave as free particles in the
cosmic plasma.

While applying a generic CSL model in cosmology, where the parameter
introduced in the theory is mass-independent, the contribution to the
total change in energy density will come from the electron fluid as
$\frac{1}{m_e}\gg\frac{1}{m_p}$ and $n_e\approx n_p$. Thus in such a
scenario the change in total energy can be written using the above
equation as
\begin{eqnarray}
\frac{\partial \varepsilon}{\partial t}=\frac{3\alpha\lambda\hbar^2}{4m_{e}}n_{e}\, .
\label{hr-const}
\end{eqnarray}
On the other hand, if one considers the variant scenario of CSL model
where the parameter $\lambda$ is dependent on mass of the particle
quadratically, then the proton fluid will contribute more to the
total change in the energy density and in such a case one can
write
\begin{eqnarray}
\frac{\partial \varepsilon}{\partial
  t}=\frac{3\alpha\lambda_0\hbar^2}{4m_{p}}n_{p}\, ,
\label{hr-prop}
\end{eqnarray}
where the mass of a nucleon can be taken as that of a proton
$m_N\approx m_p$.

It is worth pointing out at this point that the origin of the
stochastic field $W_t({\bf x)}$ in the non-linear Schr\"{o}dinger
equation of CSL model is not yet known and thus it is difficult to
point out at this moment how such a stochastic field can be
cosmologically accounted for. In this work, we will thus not consider
this stochastic field at all but only analyze the energy-increase of
standard non-relativistic particles while interacting with such
fields.

\section{Bounds on CSL parameter from Radiation dominated and Matter Dominated era}
\label{stand-cosmo}

Before investigating the spectral distortions in CMBR caused by CSL
heating of non-relativistic particles, it is important to ensure that
such heating does not hamper the standard evolution history of the
universe. We can only consider those epochs where the particles can be
treated as non-relativistic particles and hence allow one to apply
the methods of CSL model.

In standard FRW cosmology the scale factor $a$ is related to the
corresponding redshift $z$ as $1+z=\frac{a_0}{a}$ where $a_0$ is the
present scale factor. This implies that
$\frac{da}{a}=-\frac{dz}{1+z}$. It is also considered that the
universe evolves adiabatically which implies that the comoving entropy
is always conserved. For such isentropic processes $a^3T^3$ remains
constant in a comoving volume which yields $T\propto\frac1a\propto
(1+z)$ i.e. $T(z)=T_0(1+z)$ where $T_0=2.73$ K is the present
temperature of the CMBR. Writing the temperature in eV units one has
\begin{eqnarray}
T(z)\approx2.4\times10^{-4}(1+z)\,{\rm eV},
\end{eqnarray}
which implies that when the temperature of the universe is 0.5 MeV
(i.e. of the order of the rest mass energy of the electrons) the
corresponding redshift would be $2\times10^9$. Thus for radiation
dominated (RD) universe we would be interested in the redshift span of
$2\times10^9<z<3233$ and for that of matter dominated (MD) era would
be $3233<z<1$ (where $z\approx3233$ is the epoch of matter-radiation
equality).

Due to CSL heating the energy density of matter will evolve
differently than standard cosmology. According to standard cosmology
the energy density of matter gets diluted with the expansion of the
universe as
\begin{eqnarray}
\frac{d\rho_M(z)}{dt}=-3H(z)\rho_M(z),
\end{eqnarray} 
where $H\equiv\frac{\dot a}{a}$ is the Hubble parameter and we have
taken the pressure $p=0$ for matter (non-relativistic particles). But
the change in the matter energy density with time
$\left(\frac{d\rho_M}{dt}\right)$ in this case is affected by both the
expansion of the universe and the CSL heating which is proportional to
the matter number density present during that time in the cosmic
soup. Then one can write the evolution of the matter density using
Eq.~(\ref{heating-rate}), in the case where radiation and matter are
not tightly coupled, as
\begin{eqnarray}
\frac{d\rho_M(z)}{dt}=-3H(z)\rho_M(z)+\sum_{s=e,p}\frac{3\lambda\hbar^2\alpha}{4m_{s}}n_M(z).
\label{rho-dot}
\end{eqnarray}
In standard cosmology the energy density $\rho_M$ of non-relativistic
electrons and protons can be written in terms of their number density
$n_M$ as
\begin{eqnarray}
\rho_M=n_M(m_e+m_p)c^2,
\label{non-rel-stand}
\end{eqnarray}
where the number density of electron $n_e$ and that of proton $n_p$
are the same $n_e\approx n_p\approx n_M$. Using this in
Eq.~(\ref{rho-dot}) one can write as
\begin{eqnarray}
\frac{d\rho_M(z)}{\rho_M(z)}=-3\frac{da}{a}+\sum_{s=e,p}\frac{3\lambda\hbar^2\alpha}{4m_{s}(m_e+m_p)c^2}dt.
\end{eqnarray}
With the definition of Hubble parameter $H$ and the relation
$1+z=\frac{a_0}{a}$ it can be seen that 
\begin{eqnarray}
dt=-\frac{dz}{(1+z)H(z)}
\label{t-z}
\end{eqnarray}
and along with the relation $\frac{da}{a}=-\frac{dz}{1+z}$ the evolution 
equation of energy density can be written as
\begin{eqnarray}
\frac{d\rho_M(z)}{\rho_M(z)}=3\frac{dz}{1+z}-\sum_{s=e,p}\frac{3\lambda\hbar^2\alpha}{4m_{s}(m_e+m_p)c^2}\frac{dz}{(1+z)H(z)}.
\label{rho-dot1}
\end{eqnarray}
The Hubble parameter $H(z)=\frac{\dot{a}}{a}$ can be written in terms
of dimensionless quantities using the Friedmann equation as
\begin{eqnarray}
H(z)=H_0\sqrt{\Omega_{R_0}(1+z)^4 + \Omega_{M_0}(1+z)^3 + \Omega_{k_0}(1+z)^2 + \Omega_{\Lambda_0}}\,,
\end{eqnarray}
where $\Omega_i\equiv\rho_i/(3m_{\rm P}^2H^2)$ is called the density
parameter for the matter species $i$ and $\Omega_{R_0}$,
$\Omega_{M_0}$, $\Omega_{k_0}$ and $\Omega_{\Lambda_0}$ are the
present radiation, matter, curvature and dark energy density parameter
respectively. During RD era, the radiation fluid contributes dominantly
to the total density and thus the Hubble parameter can be written as 
\begin{eqnarray}
H(z)=H_0\Omega_{R_0}^{\frac12}(1+z)^2 ,
\label{rd-h}
\end{eqnarray}
whereas during MD era that would be 
\begin{eqnarray}
H(z)=H_0\Omega_{M_0}^{\frac12}(1+z)^\frac32.
\label{md-h}
\end{eqnarray}

Let us consider the RD era first. It is worth noting at this point
that during radiation era matter and radiation will be tightly coupled
and during this period Compton scattering is much more efficient than
CSL heating (as will be argued below). Also as the ratio of entropy in
baryons and photons is $10^{-9}$ the baryons will lose most of its
energy, gained by CSL heating, to photons via Compton scattering
retaining only a fraction $f$ of the order $10^{-9}$. Keeping this in
mind and using Eq.~(\ref{rd-h}) in Eq.~(\ref{rho-dot1}) one gets
\begin{eqnarray}
\frac{d\rho_M(z)}{\rho_M(z)}=3\frac{dz}{1+z}-\sum_{s=e,p}\frac{3f\lambda\hbar^2\alpha}{4m_{s}(m_e+m_p)c^2H_0\Omega_{R_0}^{\frac12}}\frac{dz}{(1+z)^3},
\end{eqnarray}
solving which one gets
\begin{eqnarray}
\rho_{M_f}=\exp\left\{K_{\rm CSL, RD}\left[\frac{1}{(1+z_f)^2}-\frac{1}{(1+z_i)^2}\right]\right\}\left(\frac{1+z_f}{1+z_i}\right)^3\rho_{M_i},
\label{rho-rd}
\end{eqnarray}
where we have defined 
\begin{eqnarray}
K_{\rm CSL,
  RD}\equiv\sum_{s=e,p}\frac{3\times10^{-9}\lambda\hbar^2\alpha}{8m_{s}(m_e+m_p)c^2\Omega_{R_0}^{\frac12}H_0}.
\end{eqnarray}
It is evident from the above equation that in absence of CSL heating
(i.e. with $\lambda=0$) we have the standard cosmological evolution of
matter density as
\begin{eqnarray}
\rho_{M_f}=\left(\frac{1+z_f}{1+z_i}\right)^3\rho_{M_i}.
\end{eqnarray}
It is important to note that before recombination the free electrons
and protons are the dominant component which contribute to the total
energy density of the universe as non-relativistic particles
i.e. $\rho_m\approx \rho_e+\rho_p$. As the number density of electron
and protons are the same, as has been considered earlier, and the
protons are much heavier than the electrons, one can consider that the
non-relativistic fluid of the cosmic plasma is dominated by the proton
fluid ($\rho_M\approx\rho_p$). Following such arguments one can ignore
the electron contributions in Eq.~(\ref{rho-rd}) and write
\begin{eqnarray}
K_{\rm CSL,
  RD}\approx\sum_{s=e,p}\frac{3\times10^{-9}\lambda\hbar^2\alpha}{8m_{s}m_pc^2\Omega_{R_0}^{\frac12}H_0}.
\end{eqnarray}
Thus to ensure that the standard evolution of the non-relativistic
plasma during RD era is not disturbed due to CSL heating one requires
\begin{eqnarray}
K_{\rm CSL,RD}\left[\frac{1}{(1+z_f)^2}-\frac{1}{(1+z_i)^2}\right]\ll1,
\end{eqnarray}
where $z_i\approx2\times10^9$ and $z_f\approx3233$.

Thus in the generic scenario of CSL model (where the total change in
energy density will depend upon the change in the electrons' energy
density) one obtains the constraint on $\lambda$ using
Eq.~(\ref{hr-const}) as
\begin{eqnarray}
\left.\lambda\right|_{\rm RD}\ll5\times10^{10}\,{\rm s}^{-1},
\end{eqnarray}
while for the variant scenario with $\lambda$ quadratically
proportional to the mass one puts an upper bound on the parameter
using Eq.~(\ref{hr-prop}) as
\begin{eqnarray}
\left.\lambda_0\right|_{\rm RD}\ll10^{14}\,{\rm s}^{-1}.
\end{eqnarray}

During MD era using Eq.~(\ref{md-h}) in Eq.~(\ref{rho-dot1}) one gets
\begin{eqnarray}
\frac{d\rho_M(z)}{\rho_M(z)}=3\frac{dz}{1+z}-\sum_{s=e,p}\frac{f\lambda\hbar^2\alpha}{4m_{s}(m_e+m_p)c^2H_0\Omega_{M_0}^{\frac12}}\frac{dz}{(1+z)^\frac52}\,,
\end{eqnarray}
and solving which yields 
\begin{eqnarray}
\rho_{M_f}=\exp\left\{K_{\rm CSL, MD}\left[\frac{1}{(1+z_f)^\frac32}-\frac{1}{(1+z_i)^\frac32}\right]\right\}\left(\frac{1+z_f}{1+z_i}\right)^3\rho_{M_i},
\label{rho-md}
\end{eqnarray}
where we have defined 
\begin{eqnarray}
K_{\rm CSL,
  MD}\equiv\sum_{s=e,p}\frac{f\lambda\hbar^2\alpha}{6m_{s}(m_e+m_p)c^2\Omega_{M_0}^{\frac12}H_0}.
\end{eqnarray}
But the situation changes in the MD era i.e. just before and after the
recombination. Here we will consider the redshift span of
$3233<z<1$. The recombination happens at $z\approx1100$ which implies
that for $z<1100$ the matter and radiation will seize to remain
tightly coupled and thereafter they will hardly interact. Thus, nearly
all the heat gained by the non-relativistic matter due to CSL heating
after $z\approx1100$ will be retained in the matter itself, yielding
$f\approx 1$. Thus the strongest bound on $\lambda$ will come from the
era $1100<z<1$. Also after recombination the energy density would be
dominated by neutral hydrogen $(\rho_M\approx\rho_H)$. Taking
$\rho_H\approx\rho_p$ we can consider $\rho_M\approx\rho_p$ throughout
$1100<z<1$. Following such arguments then one can drop the
contribution of the electrons in the MD era as well and defining
\begin{eqnarray}
K_{\rm CSL,
  MD}\approx\sum_{s=e,p}\frac{\lambda\hbar^2\alpha}{6m_{s}m_pc^2\Omega_{M_0}^{\frac12}H_0},
\end{eqnarray}
one should have
\begin{eqnarray}
K_{\rm CSL,MD}\left[\frac{1}{(1+z_f)^\frac32}-\frac{1}{(1+z_i)^\frac32}\right]\ll1,
\end{eqnarray}
to ensure that the non-relativistic fluid evolves according to the
standard cosmology during the MD era. Taking $z_i\approx1100$ and
$z_f\approx1$ for the MD era the above condition puts an upper bound
on the CSL parameter $\lambda$ for the generic case as
\begin{eqnarray}
\left.\lambda\right|_{\rm MD}\ll4\times10^{-4}\,{\rm s}^{-1}\,,
\end{eqnarray}
where we have used $\Omega_{M_0}\approx0.04$ as the present baryon
density parameter. Similarly for the mass-dependent case one gets the
upper bound as
\begin{eqnarray}
  \left.\lambda_0\right|_{\rm MD}\ll0.7\,{\rm s}^{-1}\,.
\end{eqnarray}

\section{Bounds on $\lambda$ from $\mu-$type distortion of CMB}
\label{mu-dist}

The thermalization history of the CMBR photons is affected by any
unusual energy injection during its period of thermalization with the
baryons in cosmic soup and results in spectral distortions of the
observed CMBR blackbody spectrum
\cite{Zeldovich:1969ff,Sunyaev:1970er}. The unusual energy release in
these early epochs can be due to decay of relic particles
\cite{Hu:1993gc}, by evaporation of primordial black holes
\cite{Carr:2009jm}, WIMP annihilation \cite{Khatri:2012tv}, damping of
sound waves in primordial plasma \cite{ChlubaKhatri} or by other
astrophysical mechanisms as has been mentioned in
\cite{Khatri:2012tv}. Thermalization of the photons requires
non-conservation of photon number in the cosmic soup. Cosmologically
double-Compton scattering and bremsstrahlung processes are the ones
which produce photons and thus are very significant in the
thermalization of photons in the cosmic plasma. These photon number
non-conserving processes are efficient in erasing possible spectral
distortions in the early epochs. On the other hand, the photon-number
conserving process of elastic Compton scattering,
$\gamma+e^-\rightarrow\gamma+e^-$, is the most dominant process which
couples electrons with photons at early times and redistributes the
photons in frequency. As the photon number non-conserving processes,
mainly the double Compton scattering process, become inefficient by
$z\sim 10^6$ and photon-number conserving Compton scattering process
then dominates the thermalization process, any energy release after
$z\lesssim 10^6$ leads to spectral distortion of CMBR.

Energy releases during $2\times10^6>z>5\times10^4$ lead to a kind of
CMBR spectral distortion, called $\mu-$type spectral distortion, which
yields a non-vanishing frequency dependent chemical potential of the
CMBR photons. This epoch $2\times10^6>z>5\times10^4$ is known as the
$\mu-$era in the literature. Due to such energy releases the Planck
spectrum of the thermalized photon density relaxes to a Bose-Einstein
distribution with a non-zero chemical potential $\mu$
\cite{Sunyaev:1970er} as
\begin{eqnarray}
n_{\rm Pl}\equiv\frac{1}{e^{\epsilon/\kappa_B T}-1}\quad\quad \longrightarrow \quad\quad n_{\rm BE}\equiv\frac{1}{e^{(\epsilon+\mu)/\kappa_B T}-1}  .
\end{eqnarray}
Absence of significant distortion in the CMBR spectrum from the Black
Body distribution observed by COBE/FIRAS puts an upper limit on such
distortion as $\mu\lesssim 9\times10^{-5}$ \cite{cobe}. An upcoming
space-based mission, called PIXIE, will constrain the $\mu-$type
distortion up to $\mu\sim5\times10^{-8}$ \cite{Kogut:2011xw}.

An interesting feature of these processes is that while double-Compton
scattering and bremsstrahlung processes become efficient as frequency
decreases, elastic Compton scattering process, on the other hand, is
independent of the frequency of the photons. Thus at low redshifts
when the Compton scattering is the dominant process of thermalization
of the cosmic plasma, the photon number non-conserving double-Compton
scattering and bremsstrahlung processes are still efficient at low
frequencies and are capable of retaining the black body spectrum
\cite{Hu:1992dc}. Also, the photons, which are generating at low
frequencies and getting scattered to high frequency regime, are not
sufficient to maintain a Planck spectrum at high frequencies. Thus the
elastic Compton scattering establishes a Bose-Einstein spectrum only at
the high frequencies of the spectrum.

The total $\mu-$distortion generated due to energy injections and
injection of high-energy photons during $2\times10^6>z>5\times10^4$
can be written as as \cite{Hu:1992dc}
\begin{eqnarray}
\delta\mu=\frac{1}{2.143}\left(3\frac{\delta\varepsilon}{\varepsilon}-4\frac{\delta n}{n}\right),
\label{mu}
\end{eqnarray}
where $\delta\varepsilon$ and $\delta n$ are the total changes in
energy and number density respectively with respect to a Planckian
distribution. A brief derivation of the above formula is given in
Appendix~(\ref{appn-mu}). This remarkable result shows that the total
energy distortion of $\mu-$type is independent of the actual form of
the energy injection. It can also be seen from the above equation that
direct heating of the electrons contributes in the same way as
injection of high-frequency photons as has been pointed out in
\cite{Hu:1992dc}.

It is important to note here that Compton scattering during these
epochs is very efficient to maintain the equilibrium between the
electron and photon temperature and keeps the cosmic fluid tightly
coupled. Thus the CSL heating rate should be much smaller than the
Compton scattering rate such that the thermal equilibrium of the
cosmic plasma should be retained. During these epochs equilibrium is
attained within a time scale given as \cite{Hu:1992dc}
\begin{eqnarray}
t_{\rm C}=\frac{3m_ec}{4\sigma_T\varepsilon_\gamma}\simeq \, 7.63\times10^{19}(1+z)^{-4}\,{\rm s},
\end{eqnarray}
where $\varepsilon_\gamma$ is the energy density of photons at that
redshift. Also, in the above equation the scattering of photons with
electrons is only considered as that with protons is $10^6$ times
slower and $e^-+\gamma\rightarrow e^-+\gamma$ would be the dominant
process to keep the plasma in thermal equilibrium. Thus one would
expect the CSL heating rate to be much smaller than the rate of
attaining equilibrium $(t_{\rm C}^{-1})$ via Compton scattering till
the recombination ends at $z\sim1100$, which using Eq.~(\ref{mu})
weakly puts an upper bound on the CSL heating rate for the generic
case as
\begin{eqnarray}
\left.\lambda\right|_{\rm CS}< \left.\frac{4m_e^2c^2}{3\alpha\hbar^2}\,t_{\rm C}^{-1}\right|_{z\sim1100}\approx2\times 10^3\,\rm{s}^{-1},
\end{eqnarray}
where we have used the values $m_e\approx9.1\times10^{-31}$ kg,
$\alpha\approx10^{14}$ m$^{-2}$, $c=3\times10^8$ m s$^{-1}$ and
$h=6.62\times10^{-34}$ kg m$^2$ s$^{-1}$. Similar bounds in the
variant scenario where $\lambda$ is quadratically mass-proportional
would be
\begin{eqnarray}
\left.\lambda_0\right|_{\rm CS}< \left.\frac{4m_pm_ec^2}{3\alpha\hbar^2}\,t_{\rm C}^{-1}\right|_{z\sim1100}\approx3\times 10^6\,\rm{s}^{-1}.
\end{eqnarray}

In a generic scenario of CSL model \cite{Bassi, Ghirardi},
non-relativistic particles gain energy with a rate $\lambda$ without
changing the photon number density at higher frequencies. Here we will
assume that the amount of energy gained by the electrons or protons
due to CSL heating will be transferred to the photons completely as
during these epochs the number density of photons exceeds that of the
electrons by many orders of magnitude. Thus the total $\mu-$distortion
due to CSL heating will be
\begin{eqnarray}
\delta\mu=\frac{3}{2.143}\frac{\delta\varepsilon_{\rm CSL}}{\varepsilon_\gamma},
\end{eqnarray}
where we have neglected the last term in Eq.~(\ref{mu}) as there is no
significant photon production in a generic CSL scenario at
high-frequencies. During the $\mu-$era the total amount of energy
gained by the non-relativistic fluid in the cosmic plasma can be
determined by using Eq.~(\ref{heating-rate}).  Using Eq.~(\ref{t-z})
in Eq.~(\ref{heating-rate}) yields the rate of energy gain by the
non-relativistic particles with redshift as
\begin{eqnarray}
\left.\frac{ \delta\varepsilon}{\delta z}\right.=-\sum_{s=e,p}\frac{3\alpha\lambda\hbar^2}{4m_{s}}n_{e_0}\frac{(1+z)^2}{H(z)},
\end{eqnarray}
where we have used the relation between the present number density of
electrons $n_{e_0}$ and that in a particular redshift $n_e(z)$ as
$n_e(z)=n_{e_0}(1+z)^3$. During the radiation era, i.e. before
recombination, the Hubble parameter $H(z)$ can be written as
$H(z)\simeq \Omega_{R_0}^{\frac12}H_0(1+z)^2$ where
$\Omega_{R_0}\approx6.5\times10^{-5}$ and $H_0\approx2\times10^{-18}$
s$^{-1}$ are the present radiation density parameter and Hubble
parameter respectively. Putting this value for $H(z)$ in the above
equation yields
\begin{eqnarray}
\left.\frac{\delta\varepsilon}{\delta z}\right.=-\sum_{s=e,p}\frac{3\alpha\lambda\hbar^2}{4m_{s}}\frac{n_{e_0}}{\Omega_{R_0}^{\frac12}H_0}.
\end{eqnarray}
The unperturbed Planckian photon energy at $z$ would be
\begin{eqnarray}
\varepsilon(z)=\frac{4\sigma_BT_0^4}{c}(1+z)^4,
\label{total-e}
\end{eqnarray}
which will give the fractional change in the energy of a pure
Planckian spectrum at redshift $z$ as
\begin{eqnarray}
\left.\frac{1}{\varepsilon}\frac{\delta\varepsilon}{\delta z}\right.=-\sum_{s=e,p}\frac{3\alpha\lambda\hbar^2}{16m_{s}}\frac{n_{e_0}c}{\Omega_{R_0}^{\frac12}H_0\sigma_BT_0^4}(1+z)^{-4},
\end{eqnarray} 
where $T_0\sim2.73$ K is the present temperature of the CMBR.
Integrating the above equation over $z_i=2\times10^6$ to
$z_f=5\times10^4$ one gets the total fractional energy gained by the
electron or proton plasma due to CSL heating during the $\mu-$era as
\begin{eqnarray}
\left.\frac{\Delta\varepsilon}{\varepsilon}\right.=\sum_{s=e,p}\frac{\alpha\lambda\hbar^2}{16m_{s}}\frac{n_{e_0}c}{\Omega_{R_0}^{\frac12}H_0\sigma_BT_0^4}\left[\frac{1}{(1+z_f)^3}-\frac{1}{(1+z_i)^3}\right].
\label{delta-e}
\end{eqnarray}

In a generic case, where the CSL parameter $\lambda$ is independent of
mass, we have shown that the gain in the electron energy density will
dominate and this will also be the total energy gained by the photons
via Compton scattering through electrons. Hence the total
$\mu-$distortion generated due to CSL heating will be
\begin{eqnarray}
\mu\simeq\frac{3}{2.143}\times\frac{\alpha\lambda\hbar^2}{16m_{e}}\frac{n_{e_0}c}{\Omega_{R_0}^{\frac12}H_0\sigma_BT_0^4}\left[\frac{1}{(1+z_f)^3}-\frac{1}{(1+z_i)^3}\right]\sim 1.2\times10^{-6}\lambda\,{\rm s}\,,
\end{eqnarray}
where apart from the previously mentioned values of the parameters we
have used $n_{e_0}\approx0.246$ m$^{-3}$. Using the COBE/FIRAS
measurement of $\mu-$distortion in the CMBR spectrum \cite{cobe} one
then can put an upper bound on the CSL heating parameter as 
\begin{eqnarray}
\lambda|_{\mu, \rm{COBE/FIRAS}}\lesssim 70\, {\rm s}^{-1},
\end{eqnarray}
whereas upcoming experiment PIXIE \cite{Kogut:2011xw} can put a more
stringent constraint on the CSL heating parameter as
\begin{eqnarray}
\lambda|_{\mu, \rm{PIXIE}}\lesssim 4\times10^{-2}\, {\rm s}^{-1}.
\end{eqnarray}

Similarly for the case where $\lambda$ is quadratically
mass-proportional, photons will gain energy from both the electrons
and the protons via Compton scattering. Hence, the total
$\mu-$distortion can be calculated using Eq.~(\ref{hr-prop}) and
Eq.~(\ref{total-e}) as
\begin{eqnarray}
\mu\simeq\frac{3}{2.143}\times\frac{\alpha\lambda_0\hbar^2}{16m_{p}}\times\frac{n_{e_0}c}{\Omega_{R_0}^{\frac12}H_0\sigma_BT_0^4}\left[\frac{1}{(1+z_f)^3}-\frac{1}{(1+z_i)^3}\right]\sim 6.8\times10^{-10}\lambda\,{\rm s}\,.
\end{eqnarray}
Thus bound on $\lambda_0$ from observations of COBE/FIRAS would be
\begin{eqnarray}
\lambda_0|_{\mu, \rm{COBE/FIRAS}}\lesssim 10^5\, {\rm s}^{-1},
\end{eqnarray}
and that of from PIXIE would be
\begin{eqnarray}
\lambda_0|_{\mu, \rm{PIXIE}}\lesssim 74\, {\rm s}^{-1}.
\end{eqnarray}

\section{Bounds on $\lambda$ from $y-$type distortion of CMB}
\label{y-dist}

Energy release in the early universe during $z\lesssim 5\times10^4$
leads to a $y-$type CMBR spectral distortion
\cite{Zeldovich:1969ff}. At these low redshifts Compton scattering is
not effective enough to maintain the full kinetic equilibrium between
photons and electrons. Due to energy releases if the temperature of
the electrons becomes greater than that of the photons then the
low-energy photons will be upscattered in frequency via Compton
scattering. Thus a $y-$type distortion is characterized by a deficit
of photons in the low frequency regime with an increment of photons at
high frequencies in comparison with a standard blackbody spectrum. The
efficiency of Compton scattering process to maintain a blackbody
spectrum is quantified by a parameter, called the Compton
$y-$parameter, as
\begin{eqnarray}
y=\int\frac{\kappa_b(T_e-T_\gamma)}{m_ec^2}N_e\sigma_Tcdt,
\label{y-param}
\end{eqnarray}
where $\kappa_b$ is the Boltzmann constant, $T_e$ and $T_\gamma$ are
the temperatures of the electrons and photons in the cosmic plasma
respectively and $N_e$ is the electron number density. The present
constraint on such distortion of CMBR comes from COBE/FIRAS experiments
which puts an upper bound on $y-$parameter as $y\leq1.5\times10^{-5}$
with $95\%$ confidence \cite{cobe}. The upcoming experiment PIXIE can
probe CMBR distortions up to $y\leq10^{-8}$ \cite{Kogut:2011xw}.

Energy injection of $\delta\varepsilon_\gamma$ in the photon energy
density during $z\lesssim 5\times10^4$ can yield a Compton
$y-$parameter as \cite{Zeldovich:1969ff,Chluba:2008aw}
\begin{eqnarray}
\delta y=\frac14\frac{\delta\varepsilon}{\varepsilon}.
\label{y-param1}
\end{eqnarray}
A brief derivation of the above formula is given in
Appendix~(\ref{appn-y}). In this work we are mainly interested in
$y-$type distortions arising during the pre-recombinational era
i.e. during $1100<z<5\times10^4$. This is because any energy injection
occurring after the completion of hydrogen recombination would not
imprint any additional traces in the CMBR distortions. Later on, in
post-recombinational epochs $(z\lesssim 800)$, different physical
mechanisms, like interaction of CMBR with hot intergalactic gas known
as thermal Sunyaev-Zeldovich effect \cite{Sunyaev:1972eq}; supernova
remnants at high redshifts or large-scale structure formation giving
rise to shock waves \cite{Sunyaev:1972ep,Cen:1998hc}, can also
contribute to the $y-$distortion observed in CMBR. Contribution to
$y-$distortion coming from such effects have been ignored in this
study.

As the dominant process during these epochs is still the elastic
Compton scattering, we follow the same procedure to calculate the
total energy gain by the photons, due to CSL heating of
non-relativistic electrons or protons present in the cosmic plasma, as
we have discussed while calculating the $\mu-$distortion. It is
important to note here that during the $y-$era the redshift span
$5\times10^4<z<3233$ is radiation dominated while the rest
$3233<z<1100$ is matter dominated.  Thus generalizing
Eq.~(\ref{delta-e}) and Eq.~(\ref{total-e}) to reflect this fact and
using those in Eq.~(\ref{y-param1}) in the redshift span
$z_i=5\times10^4$ to $z_f=1100$ one computes the total $y-$distortion
generated due to generic CSL heating as
\begin{eqnarray}
y&\simeq&\frac14\times\frac{3\alpha\lambda\hbar^2}{4m_{e}}\frac{n_{e_0}c}{4\sigma_BT_0^4H_0}\left[\frac{1}{3\Omega_{R_0}^{\frac12}}\left\{\frac{1}{(1+z^R_f)^3}-\frac{1}{(1+z^R_i)^3}\right\}\right.\nonumber\\
&&+\left.\frac{5}{2\Omega_{M_0}^{\frac12}}\left\{\frac{1}{(1+z^M_f)^\frac52}-\frac{1}{(1+z^M_i)^\frac52}\right\}\right]\sim 0.2\,\lambda\,\,{\rm s}\,.
\end{eqnarray}
Here we have taken $z_i^R\approx5\times10^4$ and $z_f^R\approx3233$
for the radiation dominated epoch in $y-$era and similarly
$z_i^M\approx3233$ and $z_f^M\approx1100$ for the matter dominated
epoch. Thus the constraint on $\lambda$ from the measurement of
$y-$parameter of CMBR by COBE/FIRAS experiment \cite{cobe} would be
\begin{eqnarray}
\lambda|_{y, \rm{COBE/FIRAS}}\lesssim 8\times10^{-5}\,\,{\rm s}^{-1}\,,
\end{eqnarray}
whereas in future measurement obtained by PIXIE \cite{Kogut:2011xw} can
put an upper bound on $\lambda$ as
\begin{eqnarray}
\lambda|_{y, \rm{PIXIE}}\lesssim 5\times10^{-8}\,\,{\rm s}^{-1}\,.
\end{eqnarray}

Similarly for the variant model where the parameter $\lambda$ is
mass-dependent one has for the total $y-$distortion as
\begin{eqnarray}
y&\simeq&\frac14\times\frac{3\alpha\lambda_0\hbar^2}{4m_{p}}\frac{n_{e_0}c}{4\sigma_BT_0^4H_0}\left[\frac{1}{3\Omega_{R_0}^{\frac12}}\left\{\frac{1}{(1+z^R_f)^3}-\frac{1}{(1+z^R_i)^3}\right\}\right.\nonumber\\
&&+\left.\frac{5}{2\Omega_{M_0}^{\frac12}}\left\{\frac{1}{(1+z^M_f)^\frac52}-\frac{1}{(1+z^M_i)^\frac52}\right\}\right]\sim 10^{-4}\lambda\,{\rm s}\,.
\end{eqnarray}
Thus the upper bound on $\lambda_0$ obtained from COBE/FIRAS
observations would be
\begin{eqnarray}
\lambda_0|_{y, \rm{COBE/FIRAS}}\lesssim 0.14\,\,{\rm s}^{-1}\,,
\end{eqnarray}
and that coming from PIXIE's upcoming observations
\begin{eqnarray}
\lambda_0|_{y, \rm{PIXIE}}\lesssim 10^{-4}\,\,{\rm s}^{-1}\,.
\end{eqnarray}
\section{Brief discussion on CSL model with spontaneous photon emission and its cosmological implications}
\label{photon}

In the literature it has been shown that for collapse models, atomic
systems \cite{Adler,Adler:2007kd,Fu} or free electrons \cite{Sandro}
emit radiations spontaneously when the noise term in the evolution is
treated perturbatively. The rate of creation of photons with
wavenumber $k$ due to spontaneous collapse of a (free) particle
species $s$ can be calculated as
\begin{equation}
\frac{d\Gamma(k)}{dk}\sim\frac{\lambda \alpha \hbar^2 e^2}{2 \pi^2 \epsilon_0 c^3 m_s^2k}\,. 
\label{injection}
\end{equation}
The change in the number density $n_\gamma$ of photons due to this
process would be
\begin{eqnarray}
 \left[\frac{dn_\gamma}{dt}\right]_{\rm{CSL}}\sim\frac{\lambda \alpha \hbar^2 e^2n_{s0}(1+z)^3\log{k}}{2 \pi^2 \epsilon_0 c^3 m_s^2}\approx10^{-48}\log(k)\lambda(1+z)^3,
\end{eqnarray}
where $s$ has been taken as electrons. This injection of photons into
the cosmic plasma will also contribute to the spectral distortion of
CMBR in principle, as can be seen from Eq.~(\ref{mu}). However, we can
see from the above equation that the injection of photons due to
spontaneous collapse will not provide any significant contribution at
higher frequencies and will be suppressed by $n$ as well. Even at
lower frequencies this process is clearly subdominant compared to
other photon producing processes like bremsstrahlung and double
Compton scattering where the rates of photon creation are
\cite{Hu:1992dc}
\begin{equation}
 \left[\frac{dn_\gamma}{dt}\right]_{\rm{br}}\sim \frac{10^{-18}}{k^3} \frac{\log{(2.25/x_e)}}{e^{x_e}}[1-n_\gamma(e^{x_e}-1)]z^{5/2}
\end{equation}
and
\begin{equation}
 \left[\frac{d n_\gamma}{dt}\right]_{\rm{DC}}\sim \frac{10^{-32}}{k^3} I(t)[1-n_\gamma(e^{x_e}-1)]z^5,
\end{equation}
respectively, for small $x_e$. In the above equations
\begin{eqnarray}
x_e=\frac{h ck}{k_B T_e},
\end{eqnarray}
with $T_e$ being the electron temperature and
\begin{eqnarray}
I(t)=\int dx_e\, x_e^4\,n_\gamma(n_\gamma+1).
\end{eqnarray}
We can see from above that for small $x_e$ and large $n_\gamma$
\begin{eqnarray}
\frac{\left[\frac{dn_\gamma}{dt}\right]_{\rm{CSL}}}{\left[\frac{dn_\gamma}{dt}\right]_{\rm{br}}}=
\frac{10^{-30}k^3\log(k)e^{x_e}}{\log{(2.25/x_e)}[1-n_\gamma(e^{x_e}-1)]}z^{1/2}\longrightarrow 0,
\end{eqnarray}
and also
\begin{eqnarray}
\frac{\left[\frac{dn_\gamma}{dt}\right]_{\rm{CSL}}}{\left[\frac{dn_\gamma}{dt}\right]_{\rm{DC}}}\longrightarrow 0.
\end{eqnarray}
Thus such a process at low frequencies are subdominant to other photon
creating processes like double Compton scattering and bremsstrahlung
which provide sufficient number of low energy photons to establish
blackbody spectrum at lower frequencies. However, these photon
creating processes including the one of spontaneous collapse model
become more and more inefficient as frequency of the generated photons
increases. Also the strength parameter $\lambda$ of the collapse model
is stringently constrained to very low values (as low as $10^{-5}$)
from present observations of CMBR spectral distortions. These
arguments along with the fact that the spectral distortion is
proportional to the number of photons generated which is suppressed by
the total number of photons (the factor $\frac{\delta n}{n}$ of
Eq.~(\ref{mu})) indicate that the generated photons at
high-frequencies will not be able to yield significant distortions in
the CMBR spectrum.  Also $\frac{\delta n}{n}$ in the above equation
can be significant if one takes $\lambda$ to be large which then
violates the perturbative analysis of the model. So such a photon
emitting collapse process will hardly improve the bounds on the
strength parameter $\lambda$ obtained from $\mu-$ type and $y-$ type
distortions.

\section{Discussion and Conclusion}

In the present article we have addressed the effects of CSL heating of
non-relativistic particles on standard cosmology and formation of
CMBR. As the thermal evolution of our universe has been studied
and understood rigorously by both theoretical and experimental means
during the past half a century, any anomalous heating, which can
disturb the thermal evolution of our universe, thus can be constrained
stringently by observations. Motivated by this, we endeavor to
investigate the effects a model like CSL, which heats the
non-relativistic particles with a constant rate, can have on
cosmological evolution and on CMBR formation. In the following
Table~\ref{lambda-table} we summarize the bounds we obtain by ensuring
that the CSL heating would not hamper the standard cosmological
evolution of the non-relativistic particles and also it would not
generate substantial distortions ($\mu-$type and $y-$type) in the
observed precise blackbody spectrum of CMBR.
\begin{table}[h]
\caption{Bounds on CSL strength parameter}
\centering
\begin{tabular}{c c c}
\hline\hline 
Case &$\lambda$ (in s$^{-1}$) & $\lambda_0$ (in s$^{-1}$)\\ 
\hline 
Bounds from RD era of standard cosmology & $\ll5\times10^{10}$ &$\ll10^{14}$\\ 
Bounds from MD era of standard cosmology & $\ll4\times10^{-4}$ & $\ll0.7$ \\ 
Bounds from comparing rates of Compton scattering and CSL heating &2$\times10^3$ &$3\times10^6$\\ 
Bounds from COBE/FIRAS observation of $\mu-$distortion & 70 & $10^{5}$ \\ 
Bounds from PIXIE future observation of $\mu-$distortion &$4\times10^{-2}$ & 74\\ 
Bounds from COBE/FIRAS observation of $y-$distortion &$8\times10^{-5}$ & $0.14$\\ 
Bounds from PIXIE future observation of $y-$distortion & $5\times10^{-8}$ &$10^{-4}$ \\
\hline
\end{tabular}
\label{lambda-table}
\end{table}
It is to be noted from Table~\ref{lambda-table} that the strongest
bounds are coming from the observations of $y-$distortion in the CMBR.
Such a bound on the strength parameter $\lambda$ or $\lambda_0$ is
within the detection range of the proposed diffraction experiments
from fullerene and larger molecules
\cite{Bassi:2012bg,adler-science}. It suggests that apart from
laboratory experiments, cosmological scenarios are also very important
to test any consistent collapse model which is in accordance with the
standard FRW framework of cosmology.

In \cite{Adler} a stronger bound on the strength parameter $\lambda_0$
has been obtained from the observed thermal equilibrium of IGM in the
redshift span $z\sim6-2$, by comparing the rate of heating of IGM
through CSL mechanism with the adiabatic cooling of the same due to
Hubble expansion. The slow cooling rate due to Hubble expansion
ensures that any heating mechanism during that period should also be
very small, constraining the parameter of the model to a much smaller
value (as small as $10^{-8}$ s$^{-1}$). While Adler's analysis
considers bounds yielding due to maintaining thermal equilibrium of
the IGM, the analysis done in this article studies the breakdown of
thermal equilibrium between photons and matter in cosmic plasma before
recombination due to CSL heating and thus constrains the energy intake
by the photons in the cosmic plasma before $z>1100$. As the processes,
like Compton scattering, are very efficient in redistributing the
energy over the spectrum and thus thermalizing any anomalous
heat-injection, a large enough heat injected through CSL can be
tolerated by the cosmic soup without generating large spectral
distortions and thus yielding a weaker bound on the strength parameter
than the one obtained from IGM heating \cite{Adler}. On the other
hand, while calculating the heat intake by the CMBR photons due to
heating of protons by CSL mechanism over the age of the universe, the
case which is similar to the ones considered in this article, Adler
too found a bound of $\lambda_0<10^{-5}$ s$^{-1}$ \cite{Adler} similar
to the ones obtained here.

It is nevertheless very important to mention at this point that the
above bounds have been obtained by considering only CSL effect as the
sole energy injecting process before recombination. However there can
be other mechanisms which can lead to injection of energies at such
high-redshifts like adiabatic cooling of ordinary matter
\cite{Chluba:2011hw,Khatri:2011aj}, evaporating black holes
\cite{Carr:2009jm}, decaying particles \cite{Hu:1993gc, Hu:1992dc},
dissipation of magnetic fields \cite{Jedamzik:1999bm} or
superconducting strings \cite{Ostriker:1986xc}. Some of these
processes are theoretically well understood and thus the distortion
associated with such mechanisms can be obtained precisely. Hence the
bounds on the strength parameter $\lambda$ obtained in this article
are the most conservative ones. Any other energy-release mechanism can
further tighten the bounds obtained here.  Furthermore, in this
article only white noise field has been considered for the stochastic
field $W_t(\mathbf x)$ required for the collapse processes in CSL
model. There have been attempts motivated from energy conservation to
model CSL mechanism with non-white noise \cite{NWN1,NWN2,NWN3}, which
is primarily dominant in low frequency regime. It could be interesting
to analyze this kind of collapse mechanism in the cosmological context
and can be ventured in future studies.
\appendix
\section{Brief derivations of $\mu-$type distortion}
\label{appn-mu}

Here we will give a brief derivation of $\mu-$distortion following
\cite{Hu:1992dc,Sunyaev:1970er}. For a small chemical potential
($\mu\ll1$) the energy density of a Bose-Einstein distribution is
given as
\begin{eqnarray}
\varepsilon=\frac{4\sigma_BT^4}{c}\left(1-3\frac{I_2}{I_3}\mu\right),
\label{energy-density}
\end{eqnarray}
where $\sigma_B\approx5.67\times10^{-8}$ kg s$^{-3}$ K$^{-4}$ is the
Stefan-Boltzmann constant. The number density of photons in this
distribution is
\begin{eqnarray}
n=\frac{4\sigma_BT^3}{kc}\left(1-2\frac{I_1}{I_2}\mu\right).
\label{num-density}
\end{eqnarray}
The constants $I_n$ are related to the Reimann $\zeta$ function as
\begin{eqnarray}
I_n=\int_0^\infty dx\frac{x^n}{e^x-1}=n!\zeta(n+1),
\end{eqnarray}
with the numerical values $I_1\simeq1.645$, $I_2\simeq2.404$ and
$I_3\simeq6.494$. Energy release processes in the early epochs may
involve direct energy injection or injection of high frequency
photons. The change in the energy density given in
Eq.~(\ref{energy-density}) is given by
\begin{eqnarray}
\frac{\delta\varepsilon}{\varepsilon}=4\,\delta\ln T-3\frac{I_2}{I_3}\delta\mu,
\end{eqnarray}
whereas the change in the number density given in
Eq.~(\ref{num-density}) will be
\begin{eqnarray}
\frac{\delta n}{n}=3\,\delta\ln T-2\frac{I_1}{I_2}\delta\mu.
\end{eqnarray}
Solving the above two equations simultaneously one obtains the change
in chemical potential of the photons in the Bose-Einstein distribution
as given in Eq.~(\ref{mu}).

\section{Brief derivations of $y-$type distortion}
\label{appn-y}

Here we will give a brief derivation to calculate the $y-$parameter
generated due to energy injection. In the non-relativistic limit,
described by $\kappa_bT_e\ll m_ec^2$ and $\kappa_bT_\gamma\ll m_ec^2$,
Kompaneets' equation is the kinetic equation of Compton scattering
which can be expressed as \cite{Hu:1992dc}
\begin{eqnarray}
\frac{\partial n_\gamma}{\partial t}=N_e\sigma_Tc\left(\frac{\kappa_bT_e}{m_ec^2}\right)\frac{1}{x_e^2}\frac{\partial}{\partial x_e}\left[x_e^4\left(\frac{\partial n_\gamma}{\partial x_e}+n_\gamma+n_\gamma^2\right)\right],
\end{eqnarray}
where $n_\gamma$ is the photon occupation number and
$x_e\equiv\frac{h\nu}{\kappa_bT_e}$. The above equation simplifies
when $T_e$ is much larger, yielding \cite{Zeldovich:1969ff}
\begin{eqnarray}
\frac{\partial n_\gamma}{\partial t}=N_e\sigma_Tc\left(\frac{\kappa_bT_e}{m_ec^2}\right)\frac{1}{x_e^2}\frac{\partial}{\partial x_e}\left(x_e^4\frac{\partial n_\gamma}{\partial x_e}\right).
\end{eqnarray}
Using the definition of Compton $y-$parameter given in
Eq.~(\ref{y-param}) the above equation can be written as 
\begin{eqnarray}
\frac{\partial n_\gamma}{\partial y}=\frac{1}{x_e^2}\frac{\partial}{\partial x_e}\left(x_e^4\frac{\partial n_\gamma}{\partial x_e}\right),
\label{num-eq}
\end{eqnarray}
where we have dropped the term $T_\gamma$ in Eq.~(\ref{y-param}) as in
this case $T_e>T_\gamma$. The energy density $\varepsilon_\gamma$ of
photons can be written in terms of occupation number $n_\gamma$ as
\begin{eqnarray}
\varepsilon_\gamma=\frac{8\pi h}{c^3}\int n_\gamma\nu^3d\nu=\frac{8\pi}{h^3c^3}\left(\kappa_bT_e\right)^4\int n_\gamma x_e^3dx_e.
\end{eqnarray}
Thus multiplying both sides of Eq.~(\ref{num-eq}) by $x_e^3$ and
solving for the integration yields
\begin{eqnarray}
\frac{\delta \varepsilon_\gamma}{\delta y}=4\varepsilon_\gamma.
\end{eqnarray}
Thus energy injection of $\delta\varepsilon_\gamma$ during $z\lesssim
5\times10^4$ can yield a Compton $y-$parameter as given in
Eq.~(\ref{y-param1}).
\begin{acknowledgements}

This article is made possible through the support of a grant (ID:
20768) from the John Templeton Foundation. The opinions expressed in
this publication are those of the authors and do not necessarily
reflect the views of the John Templeton Foundation.  KL and SD wish to
thank Satyabrata Sahu, Cenalo Vaz and Tirthankar Roy Choudhury for
helpful discussions. AB acknowledges partial support from MIUR (PRIN
2008), INFN and the COST Action MP1006 ``Fundamental Problems in
Quantum Physics". The authors also thank Stephen Adler, Rishi Khatri
and T. P. Singh for carefully reading the draft and giving useful
comments.

\end{acknowledgements}


\begin{thebibliography}{99}
\bibitem{Bassi}
 A.~Bassi and G.~C.~Ghirardi,
  Phys.\ Rept.\  {\bf 379}, 257 (2003)
  [arXiv:quant-ph/0302164].
\bibitem{Bassi:2012bg} 
  A.~Bassi, K.~Lochan, S.~Satin, T.~P.~Singh and H.~Ulbricht,
  arXiv:1204.4325 [quant-ph].
\bibitem{Gisin}
 N.~Gisin,
  Helv.\ Phys.\ Acta {\bf 62}, 363 (1989).
\bibitem{Ghirardi}
  G.~C.~Ghirardi, A.~Rimini and T.~Weber,
  Phys.\ Rev.\  D {\bf 34}, 470 (1986).
\bibitem{Diosi}
L.~Di\'{o}si,  Phys.\ Rev.\  A {\bf 40}, 1165 (1989).
\bibitem{Bassi:qmupl} 
  A.~Bassi, E.~Ippoliti and B.~Vacchini,
 J.\ Phys.\  A: Math.\ Gen.\ {\bf 38}, 8017 (2005)
  [arXiv:quant-ph/0506083].
\bibitem{Ghirardi:1989cn} 
  G.~C.~Ghirardi, P.~M.~Pearle and A.~Rimini,
  Phys.\ Rev.\ A {\bf 42}, 78 (1990).
\bibitem{PS} 
P.~Pearle and E.~Squires, Phys.\ Rev.\ Lett.\ {\bf 73} (1994).
\bibitem{Karolyhazy}
F.~Karolyhazy, A.~Frenkel and B.~Lukacs, ``{\it Physics as natural philosophy}, '' (1982) , Edited by : A.~Shimony and H.~Feshbach (MIT Press, Cambridge).
\bibitem{Diosi-gr}
L.~Di\'{o}si,  Phys.\ Letts.\  A {\bf 120}, 377 (1987).
\bibitem{penrose}
R.~Penrose,  Gen.\ Rel.\  Grav {\bf 28}, 581 (1996).
\bibitem{adler-science}
S.~Adler and A.~Bassi, Science\ {\bf 325} 275 (2009)
\bibitem{Adler}
S.~L.~Adler,
  J.\ Phys.\ A A {\bf 40}, 2935 (2007)
  [Erratum-ibid.\ A {\bf 40}, 13501 (2007)]
  [quant-ph/0605072].
\bibitem{Sunyaev:1970er}
  R.~A.~Sunyaev and Y.~B.~Zeldovich,
  Astrophys.\ Space Sci.\  {\bf 7}, 20 (1970).
\bibitem{cobe}
J.~C.~Mather {\it et al.},
  Astrophys.\ J.\  {\bf 420}, 439 (1994).
\\
D.~J.~Fixsen, E.~S.~Cheng, J.~M.~Gales, J.~C.~Mather, R.~A.~Shafer and E.~L.~Wright,
  Astrophys.\ J.\  {\bf 473}, 576 (1996)
  [arXiv:astro-ph/9605054].
\bibitem{Kogut:2011xw}
  A.~Kogut {\it et al.},
  JCAP {\bf 1107}, 025 (2011)
  [arXiv:1105.2044 [astro-ph.CO]].
\bibitem{Zeldovich:1969ff}
  Y.~B.~Zeldovich and R.~A.~Sunyaev,
  Astrophys.\ Space Sci.\  {\bf 4}, 301 (1969).
\bibitem{Hu:1993gc}
  W.~Hu and J.~Silk,
  Phys.\ Rev.\ Lett.\  {\bf 70}, 2661 (1993).
\bibitem{Carr:2009jm}
  B.~J.~Carr, K.~Kohri, Y.~Sendouda and J.~Yokoyama,
  Phys.\ Rev.\  D {\bf 81}, 104019 (2010)
  [arXiv:0912.5297 [astro-ph.CO]].
\bibitem{Khatri:2012tv} 
  R.~Khatri and R.~A.~Sunyaev,
  arXiv:1203.2601 [astro-ph.CO].
\bibitem{ChlubaKhatri}
J.~Chluba, R.~Khatri and R.~A.~Sunyaev,
arXiv:1202.0057 [astro-ph.CO].
\bibitem{Hu:1992dc}
  W.~Hu and J.~Silk,
  Phys.\ Rev.\  D {\bf 48}, 485 (1993).
\bibitem{Chluba:2008aw}
  J.~Chluba and R.~A.~Sunyaev,
  arXiv:0803.3584 [astro-ph].
\bibitem{Sunyaev:1972eq}
  R.~A.~Sunyaev and Y.~B.~Zeldovich,
  Comments Astrophys.\  Space Phys.\  {\bf 4}, 173 (1972).
\bibitem{Sunyaev:1972ep}
  R.~A.~Sunyaev and Y.~B.~Zeldovich,
  Astron.\ Astrophys.\  {\bf 20}, 189 (1972).
\bibitem{Cen:1998hc}
  R.~Cen and J.~P.~Ostriker,
  Astrophys.\ J.\  {\bf 514}, 1 (1999)
  [arXiv:astro-ph/9806281].
\bibitem{Adler:2007kd} 
  S.~L.~Adler and F.~M.~Ramazanoglu,
  J.\ Phys.\ A A {\bf 40}, 13395 (2007)
  [J.\ Phys.\ A A {\bf 42}, 109801 (2009)]
  [arXiv:0707.3134 [quant-ph]].
\bibitem{Fu} Q.~Fu, Phys.\ Rev.\ A {\bf 56} 1806 (1997).
\bibitem{Sandro}
S.~L.~Adler, A.~Bassi and S.~Donadi,
[arXiv:1011.3941 [quant-ph]].
\bibitem{Chluba:2011hw} 
  J.~Chluba and R.~A.~Sunyaev,
  arXiv:1109.6552 [astro-ph.CO].
\bibitem{Khatri:2011aj} 
  R.~Khatri, R.~A.~Sunyaev and J.~Chluba,
  Astron.\ Astrophys.\  {\bf 540}, A124 (2012)
  [arXiv:1110.0475 [astro-ph.CO]].
\bibitem{Jedamzik:1999bm} 
  K.~Jedamzik, V.~Katalinic and A.~V.~Olinto,
  Phys.\ Rev.\ Lett.\  {\bf 85}, 700 (2000)
  [astro-ph/9911100].
\bibitem{Ostriker:1986xc} 
  J.~P.~Ostriker, A.~C.~Thompson and E.~Witten,
  Phys.\ Lett.\ B {\bf 180}, 231 (1986).
\bibitem{NWN1} 
S.~L.~Adler and A. Bassi,
J. Phys. A: Math. Theor.{\bf 40} (2007) 15083-15098.
\bibitem{NWN2} 
S.~L.~Adler and A. Bassi,
J. Phys. A: Math. Theor.{\bf 41} (2008) 395308.
\bibitem{NWN3} 
L. Ferialdi and A. Bassi,
Phys. \ Rev. \ Lett. {\bf 108}, 170404 (2012).
\end{thebibliography}
\end{document}